\documentclass[aps,pre,reprint,superscriptaddress,longbibliography]{revtex4-2}

\usepackage[utf8]{inputenc}
\usepackage[T1]{fontenc}
\usepackage{graphicx}
\usepackage{amsmath,amssymb}
\usepackage{hyperref}
\usepackage{xcolor}
\hypersetup{
    colorlinks=true,
    linkcolor=blue,
    citecolor=blue,
    urlcolor=blue
}
\usepackage{tensor}
\usepackage{textgreek}
\usepackage{bm}
\usepackage{mathrsfs}

\makeatletter
\renewcommand{\@fnsymbol}[1]{%
  \ifcase#1\or †\or ‡\or *\or **\or \ddagger\or \mathsection\or \mathparagraph\or \|\or **\else\@arabic{#1}\fi}
\makeatother

\begin{document}

\title{Emergent Dynamics of Active Systems on Curved Environments}

\author{Euan D. Mackay}
\thanks{E.D.M.\ and G.J.\ contributed equally to this work.}
\affiliation{School of Life Sciences, University of Dundee, Dundee, DD1 5EH, United Kingdom}

\author{Giulia Janzen}
\thanks{E.D.M.\ and G.J.\ contributed equally to this work.}
\affiliation{Department of Theoretical Physics, Complutense University of Madrid, Madrid, 28040, Spain}

\author{D. A. Matoz-Fernandez}
\email{dmatoz@ucm.es}
\affiliation{Department of Theoretical Physics, Complutense University of Madrid, Madrid, 28040, Spain}

\author{Rastko Sknepnek}
\email{r.sknepnek@dundee.ac.uk}
\affiliation{School of Life Sciences, University of Dundee, Dundee, DD1 5EH, United Kingdom}
\affiliation{School of Science and Engineering, University of Dundee, Dundee, DD1 4HN, United Kingdom}

\date{\today}

\begin{abstract}
Curvature plays a central role in the proper function of many biological processes. With active matter being a standard framework for understanding many aspects of the physics of life, it is natural to ask what effect curvature has on the collective behaviour of active matter. In this paper, we use the classical theory of surfaces to explore the active motion of self-propelled agents confined to move on a smooth curved two-dimensional surface embedded in Euclidean space. Even without interactions and alignment, the motion is non-trivially affected by the presence of curvature, leading to effects akin, e.g.\ to gravitational lensing and tidal forces. Such effects can lead to intermittent trapping of particles and profoundly affect their flocking behaviour. We show that these effects are governed by a geometric torque that, in the absence of noise and interactions, compels particles to move along geodesics.  
\end{abstract}

\maketitle

%---------------- Main Text ----------------%
\section{Introduction}
Active matter \cite{ramaswamy2010mechanics, vicsek2012collective,MarchettiRev} has emerged as a powerful framework for understanding the physics of life \cite{needleman2017active,lauga2020fluid,pismen2021active,kurzthaler2023isactive}, describing the dynamics, organisation, and collective behaviour of active agents such as cells. The same principles apply to non-living active systems, highlighting that a defining feature of active matter is the continuous conversion of stored or ambient free energy by each agent into mechanical work, whether as self-propulsion or internal stresses. This continuous, localised energy input drives the system far from equilibrium, giving rise to phenomena with no direct counterparts in equilibrium physics. As a result, active systems can display a wide range of collective behaviours, including large-scale flocking~\cite{Vicsek1995, toner2024physics}, motility-induced phase separation \cite{Fily2012,cates2015motility}, active turbulence~\cite{Narayan2007, Theurkauff2012, Palacci2013,yeomans2023active}, etc. 

Biological systems, however, are typically not flat, with curvature appearing across a wide range of length scales, from the subcellular structures and individual cells to tissues and organs. The presence of curvature is tightly connected to biological function \cite{schamberger2023curvature}. For example, mitochondria undergo shape change before triggering cell death \cite{cereghetti2006many}. Corneal cells rely on hemispherical geometry to migrate in a spiral pattern, which is important for maintaining homeostasis while constantly regenerating the tissue \cite{kostanjevec2025spiral}. During embryonic development, initially flat sheets of cells fold and bend into complex three-dimensional forms such as branches, tubes, and furrows, with cellular organisation and fate heavily influenced by local curvature \cite{Chuai2006, Buske2011, Callens2020substrate}. 

In vitro systems, such as microtubule-dynein mixtures confined to move on the surface of spherical \cite{Keber2014} and toroidal \cite{ellis2018curvature} vesicles, have revealed a fascinating interplay between the active dynamics of topological defects in the nematic order and the underlying curvature. The interplay between activity, geometry, and deformability of vesicles gives rise to a tunable periodic state in which topological defects oscillate between tetrahedral and planar configurations, or even drive shape changes such as the formation of protruding jets \cite{Keber2014}. When confined to ellipsoidal surfaces, active nematics exhibit similar but more complex oscillatory behaviour, with non-uniform curvature introducing richer defect dynamics \cite{clairand2024dynamics}. In contrast, toroidal geometries lead to activity-driven unbinding of oppositely charged topological defects, which segregate into regions of opposite Gaussian curvature \cite{ellis2018curvature}. Similarly, a mixture of actin filaments and myosin motors confined to the surface of a spherical vesicle showed that spherical confinement can shape activity-induced patterns \cite{Hsu2022}. Spheroids of migrating human mammary cells exhibit curvature-induced propagating single-wavelength velocity waves \cite{brandstatter2023curvature}.

Inspired by these phenomena, a growing number of studies explored the effects arising from the coupling between activity, geometry, and topology. Those studies included models that integrate active matter dynamics with viscoelastic theory of shells and sheets \cite{berthoumieux2014active,metselaar2019topology,salbreux2017mechanics, matoz2020wrinkle, mietke2019self, khoromskaia2023active}, explored the role of the dynamics of topological defects during morphogenesis \cite{hoffmann2022theory,vafa2022active,wang2023patterning}, used agent-based models to describe the effects of curvature on collective active dynamics \cite{Sknepnek2015,Henkes2018,ehrig2017curvature,bruss2017curvature, castro2018active,schonhofer2022curvature,nemeth2025intrinsic}, explored actively driven flows on curved surfaces \cite{mickelin2018anomalous,napoli2020spontaneous,rank2021active}, and investigated the optimal navigation strategies on manifolds \cite{Piro2021}.

An emerging picture is that curvature can non-trivially affect collective active behaviours, often giving rise to striking motion patterns, for example, the rotating band state \cite{Keber2014, Sknepnek2015, Hsu2022}. Given the fundamental role of curvature in many biological systems, there has been a growing interest in understanding its influence in active matter systems. Before applying curved active-matter concepts to specific biological systems, it is useful to first establish a framework using minimal examples. Such toy models highlight the effects of curvature itself without obscuring the discussion with additional phenomena present in real systems.

In this work, we discuss a theoretical framework to describe active agents confined to curved surfaces, drawing from the classical theory of surfaces embedded in $\mathbb{R}^3$. We demonstrate how curvature enters the equations of motion and affects agent dynamics, giving rise to effects reminiscent of gravitational lensing and tidal forces. By adopting this simplified, yet general approach, we isolate the core geometric contributions to active behaviour and establish a clear foundation for interpreting more complex collective phenomena. In particular, we introduce the notion of a geometric torque that, in the absence of noise and interactions, directs particles along geodesics. This concept also provides a natural explanation for emergent behaviours such as the band formation exhibited by softly repulsive active agents on a spherical surface \cite{Sknepnek2015}. 

The paper is organised as follows. In Section \ref{sec:model}, we present the equations of motion for active Brownian particles confined to move on a smooth curved surface, with particular emphasis on the role of curvature. In Section \ref{sec:results}, we apply the model to several simple cases that showcase the effects of curvature on the motion of a single active particle, a flock both with and without interactions, and a band-state of softly-repulsive particles on a sphere. In Section \ref{sec:discussion}, we summarise our main findings and provide an outlook for further investigations. Finally, in the appendices, we summarise key concepts of the theory of surfaces. 

\section{Model}
\label{sec:model}

We first consider active agents (i.e.\ self-propelled particles) confined to move on the $xy$ plane in $\mathbb{R}^3$. Each agent is propelled in the direction of a unit vector $\mathbf{n}_i$ by an active force of constant magnitude $f_0$. Hydrodynamic interactions are neglected throughout, i.e.\ we assume the dry limit \cite{MarchettiRev}. We can write $\mathbf{n}_i=\cos\vartheta_i\mathbf{e}_\mathrm{x} + \sin\vartheta_i\mathbf{e}_\mathrm{y}$, where $\mathbf{e}_\mathrm{x}$ and $\mathbf{e}_\mathrm{y}$ form the orthonormal basis in the plane. Since the plane is flat, we can use the same global basis to describe the orientation of each $\mathbf{n}_i$. If $\mathbf{r}_i$ denotes the position of the particle $i$ at time $t$, in the overdamped limit, the equations of motion for $\mathbf{r}_i$ and $\vartheta_i$ are \cite{szabo2006phase, henkes2011active, baconnier2025self},
\begin{subequations}
    \begin{eqnarray}
        \gamma_\mathrm{t}\dot{\mathbf{r}}_i &=& f_0\mathbf{n}_i + \mathbf{F}_i + \boldsymbol{\xi}_i(t),\\
        \gamma_\mathrm{r}\dot{\vartheta}_i &=& \tau_i + \eta_i(t), \label{eq:theta-plane}
    \end{eqnarray}
\end{subequations}
where $\gamma_\mathrm{t}$ and $\gamma_\mathrm{r}$ are, respectively, the translational and rotational friction coefficients, $\mathbf{F}_i$ is the total passive force acting on the particle, and $\tau_i$ is the ``scalar'' torque (i.e.\ the torque $\boldsymbol{\tau}_i=\tau_i\mathbf{e}_\mathrm{z}$) acting to rotate $\mathbf{n}_i$ around the $z-$axis. Moreover, $\boldsymbol{\xi}_i$ is the delta-correlated random force with $\langle\boldsymbol{\xi}_i(t)\rangle=0$ and $\langle\xi_{i,\alpha}(t)\xi_{j,\beta}(t')\rangle = 2D_\mathrm{t}\delta_{ij}\delta_{\alpha\beta}\delta(t-t')$, where $\alpha,\beta\in\{\mathrm{x},\mathrm{y}\}$, $D_\mathrm{t}$ is the translational diffusion constant, and $\langle\dots\rangle$ denotes the average over realisations of the noise. Similarly, $\eta_i(t)$ is the random torque with $\langle\eta_i(t)\rangle=0$ and $\langle\eta_i(t)\eta_j(t')=2D_\mathrm{r}\delta_{ij}\delta(t-t')$, with $D_\mathrm{r}$ being the rotational diffusion coefficient. Note that we do not explicitly relate the diffusion and friction coefficients, as the system is active and the fluctuation-dissipation theorem does not generally hold.

\begin{figure}[t]
\includegraphics[width=1\columnwidth]{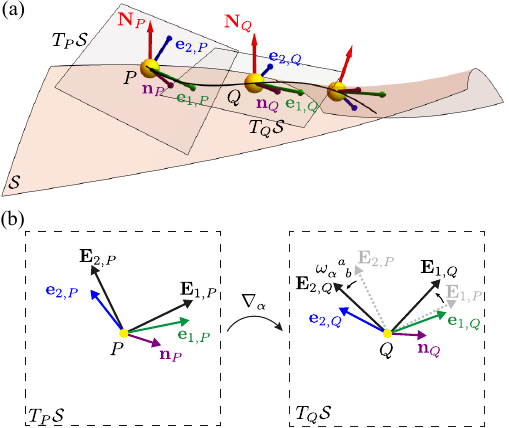}
\caption{\label{fig:model} (a) Active agents (yellow) are confined to move on the surface $\mathcal{S}$ (pale orange) embedded in $\mathbb{R}^3$. Each agent is endowed with a unit-length vector $\mathbf{n}_i$ (purple) tangent to $\mathcal{S}$ at every point; it sets the direction of the self-propulsion. Tangent planes at every point are spanned by a pair of tangent vectors defined as $\mathbf{e}_\alpha=\partial_\alpha\mathbf{r}$ (green and blue). We use subscripts $P$  and $Q$ to denote that the tangent vectors depend on the point on the surface. Vector $\mathbf{N}$ (red) is the unit normal to the surface, which also depends on the position. (b) The coordinate basis $\{\mathbf{e}_1,\mathbf{e}_2\}$ (green and blue) in the tangent space $T_P\mathcal{S}$ is not necessarily orthonormal. One can, however, choose an orthonormal (i.e.\ frame) basis $\{\mathbf{E}_1,\mathbf{E}_2\}$ (black). Vectors in tangent spaces $T_P\mathcal{S}$ and $T_P\mathcal{S}$ are related via Levi-Civita connection ($\nabla_\alpha$). Expressed in the frame bases, connection coefficients are known as the spin connection ($\tensor{\omega}{_\alpha^a_b}$). It acts to rotate the frame basis around the surface normal as one moves along $\mathcal{S}$.  }
\end{figure}

We now consider the curved case, where active agents are constrained to move on a curved surface embedded in three-dimensional Euclidean space. We do not discuss the mechanism that enforces this constraint, which depends on the specific experimental realisation, but assume that the surface acts as a time-independent holonomic constraint. In the presence of the constraint, the description active motion is more subtle. Let the surface $\mathcal{S}$ be described by a vector $\mathbf{r}(u^1,u^2)\in\mathbb{R}^3$ parametrised by curvilinear coordinates $u^1$ and $u^2$ with $u^1,u^2\in D\subseteq\mathbb{R}^2$ (Appendix \ref{app:A}). Two tangent vectors $\mathbf{e}_\alpha = \frac{\partial\mathbf{r}}{\partial u^\alpha}\equiv\partial_\alpha\mathbf{r}$, with $\alpha\in\{1,2\}$, form the so-called coordinate basis of the tangent plane at a point. Tangent planes at different points are, however, not necessarily parallel in $\mathbb{R}^3$ (Fig.\ \ref{fig:model}a), and relating vectors in different planes becomes non-trivial, leading to interesting effects on the active dynamics.  

The trajectory of agent $i$ is a curve $\boldsymbol{\gamma}_i(t)=\mathbf{r}_i(u^1(t),u^2(t))$ embedded in $\mathcal{S}$ and parametrised by time (Appendix \ref{app:B}). The velocity vector is tangent to the curve and, in the local coordinate basis,   $\mathbf{v}_i=\dot{\mathbf{r}}_i=\dot{u}^\alpha\partial_\alpha\mathbf{r}_i=\dot{u}^\alpha\mathbf{e}_\alpha$, where the overdot indicates the time derivative and we assume summation over pairs of repeated indices. Vectors $\mathbf{e}_\alpha$ should also carry the label $i$ to indicate that they belong to the tangent space at the position $\mathbf{r}_i$. However, we omit it here to reduce clutter. In curvilinear coordinates, the equation of motion for the agent's position becomes
\begin{equation}
    \gamma_\mathrm{t}\dot{u}_i^\alpha = f_0 n_i^\alpha + \tilde{F}_i^\alpha + \xi_i^\alpha. 
    \label{eq:r-curved}
\end{equation}
Vector $\mathbf{n}_i$ is already in the tangent plane. The random force $\boldsymbol{\xi}_i$ modelling the noise can also be assumed to act in the tangent plane, which leads to some subtleties \cite{elworthy1998stochastic,del2017stochastic}. For simplicity, however, we neglect the noise. Furthermore, the passive force $\mathbf{F}_i$, is typically a result of short-range pairwise interactions, i.e.\ $\mathbf{F}_i=\sum_{j\in\mathcal{N}_i}\mathbf{f}_{ij}$, where $\mathcal{N}_i$ is a neighbourhood of the agent $i$ and $\mathbf{f}_{ij}$ is the pairwise interaction between agents $i$ and $j$. In many cases, $\mathbf{f}_{ij}$ is not necessarily constrained by the surface, and $\mathbf{F}_i$ is not expected to point in the tangent direction. Therefore, to keep the constraint satisfied, we need first to project it to the tangent plane to obtain $\tilde{\mathbf{F}}_i=\mathbf{F}_i - \left(\mathbf{F}_i\cdot\mathbf{N}_i\right)\mathbf{N}_i$, where $\mathbf{N}_i$ is the local unit normal (Eq.\ \eqref{eq:surface_normal} in Appendix \ref{app:A}). We then have $\tilde{\mathbf{F}}_i=\tilde{F}_i^\alpha\mathbf{e}_\alpha$. Note that all dot products are in the embedding Euclidean space. Finally, Eq.\ (\ref{eq:r-curved}) is typically non-dimensionalised by constructing the appropriate units of length, time, force, etc.

Assuming that the appropriate non-dimensionalisation has been carried out, the dynamics of vector $\mathbf{n}_i$ involves it being both rotated and transported along the trajectory, i.e.
\begin{equation}
    \label{eq:n-curved}
    \frac{\mathrm{D}\mathbf{n}_i}{\mathrm{d}t} = \tilde{\boldsymbol{\tau}}_i\times\mathbf{n}_i + \eta_i(t)\mathbf{N}_i\times\mathbf{n}_i,
\end{equation}
where $\tilde{\boldsymbol{\tau}}_i=\left(\boldsymbol{\tau}_i\cdot\mathbf{N}_i\right)\mathbf{N}_i$ is the projection of the torque vector $\boldsymbol{\tau}_i$ onto the local surface normal $\mathbf{N}_i$ and $\eta_i(t)\mathbf{N}_i\times\mathbf{n}_i$ is the torque due to noise, which we neglect. Vector $\mathbf{N}_i\times\mathbf{n}_i$ is in the tangent plane and is orthogonal to $\mathbf{n}_i$, reflecting the fact that the first term in Eq.\ \eqref{eq:n-curved} acts to rotate $\mathbf{n}_i$ around the local surface normal. The reason for projecting the torque $\boldsymbol{\tau}_i$ onto $\mathbf{N}_i$ is analogous to why we projected the force onto the tangent plane, i.e.\ to ensure that active agents are confined to the surface. The derivative of the vector $\mathbf{n}_i$ along the curve with local tangent vector $\mathbf{v}_i$ is given by $\frac{\mathrm{D}\mathbf{n}_i}{\mathrm{d}t}=\nabla_{\mathbf{v}_i}\mathbf{n}_i = \dot{u}^\alpha\nabla_\alpha\mathbf{n}_i=\left(\dot{u}^\alpha\partial_\alpha n_i^\beta + \tensor{\Gamma}{^\beta_{\alpha\gamma}}\dot{u}^\alpha n_i^\gamma\right)\mathbf{e}_\beta$. Here, $\nabla_\alpha$ denotes the covariant derivative and $\tensor{\Gamma}{^\beta_{\alpha\gamma}}$ are the associated Christoffel symbols (Appendix \ref{app:B}). Using the chain rule for partial derivatives, we can write $\dot{u}^\alpha\partial_\alpha n_i^\beta = \dot{n}_i^\beta$. After rearranging terms and relabelling indices, we obtain the equation of motion for the components of $\mathbf{n}_i$ in the coordinate basis,
\begin{equation}
    \dot{n}_i^\alpha = \left(\boldsymbol{\tau}_i\cdot\mathbf{N}_i\right)\left(\mathbf{N}_i\times\mathbf{n}_i\right)^\alpha - \tensor{\Gamma}{^\alpha_{\beta\gamma}}\dot{u}^\beta n_i^\gamma. \label{eq:n-dyn-coordinate}
\end{equation}
The last term is purely geometric, reflecting that the agents are being transported along a curve embedded in a surface. In other words, it ensures that vector $\mathbf{n}_i$ remains tangent to the surface as it is transported along it. In the planar case, all Cristofell symbols are equal to zero and $\mathbf{N}_i=\mathbf{e}_\mathrm{z}$ everywhere, and one readily recovers Eq.\ \eqref{eq:theta-plane}. In the absence of all passive forces and torques and noise, $\mathbf{v}_i = v_0\mathbf{n}_i$ with $v_0=f_0/\gamma_\mathrm{t}$. Under these conditions, the first term on the right-hand side in Eq.\ \eqref{eq:n-dyn-coordinate} is zero. Moving the second term to the left and using $\dot{u}_i^\beta=v_0n_i^\beta$, Eq.\ \eqref{eq:n-dyn-coordinate} reduces to 
\begin{eqnarray}
\dot{n}_i^\alpha+v_0\tensor{\Gamma}{^\alpha_{\beta\gamma}}n_i^\beta n_i^\gamma&=&v_0\left(n_i^\beta\partial_\beta n_i^\alpha+\tensor{\Gamma}{^\alpha_{\beta\gamma}}n_i^\beta n_i^\gamma\right)\nonumber\\
&=&0.
\end{eqnarray}
The expression in the brackets is the $\alpha$ component of the vector $\nabla_{\mathbf{n}_i}\mathbf{n}_i$, which defines the geodesic curvature of a curve with tangent vector $\mathbf{n}_i$. Therefore, we have $\nabla_{\mathbf{n}_i}\mathbf{n}_i=0$, i.e.\ a non-interacting active particle follows a geodesic (Appendix \ref{app:B}). The geodesic is completely determined (up to trivial reparametrisation) by the particle's initial position and initial tangent vector.

In general, the coordinate tangent basis is not orthonormal. However, it is possible to select an orthonormal basis, e.g.\ using the Gram-Schmidt procedure. This basis is known as the local frame, and we denote it as $\{\mathbf{E}_1,\mathbf{E}_2\}$. The local frame is not associated with any specific coordinates, i.e.\ it does not arise from a surface parametrisation. We use Latin indices from the beginning of the alphabet (e.g.\ $a$, $b$, $c$, etc.) to label its elements, i.e.\  $\mathbf{E}_a$ with $a\in\{1,2\}$. We also assume summation over pairs of repeated Latin indices. The coordinate and frame bases are related to each other by a linear transformation $\mathbf{e}_\alpha=\tensor{\Lambda}{^a_\alpha}\mathbf{E}_a$, where $\tensor{\Lambda}{^a_\alpha}$ are elements of the tetrad or vielbein. In the case of a two-dimensional surface, the tetrad is known as the zweibein (Appendix \ref{app:C}) \cite{kamien2002geometry,bowick2009two,wald2010general}.  In the frame basis, the vector $\mathbf{n}_i$ can be written as $\mathbf{n}_i = n^a_i\mathbf{E}_a$. The covariant derivative is then $\nabla_\alpha\mathbf{n}_i = \left(\partial_\alpha n^a_i + \tensor{\omega}{_\alpha^a_b}n^b_i\right)\mathbf{E}_a$, and the directional derivatives is $\nabla_{\mathbf{v}}\mathbf{n}_i = \left(\dot{u}^\alpha\partial_\alpha n_i^a + \tensor{\omega}{_\alpha^a_b}\dot{u}^\alpha n_i^b\right)\mathbf{E}_a$, with $\mathbf{v}=\dot{u}^\alpha\mathbf{e}_\alpha$. The terms $\omega_{\alpha ab} = \tensor{\omega}{_\alpha^c_b}\delta_{ac}$ are components of the spin connection, which encodes the Levi-Civita connection \cite{lee2018introduction} in the frame basis (Fig.\ \ref{fig:model}b). They are the analogue of the Christoffel symbols, but expressed in the frame rather than in the coordinate basis. The Christoffel symbols and the spin connection are linked by the zweibeins (Appendix \ref{app:C}). We used $\delta_{ab}$ to lower one of the indices, since in the frame basis the components of the metric tensor are given by $g_{ab}=\mathbf{E}_a\cdot\mathbf{E}_b=\delta_{ab}$ (Appendix \ref{app:C}). 

In the case of a two-dimensional surface, $\omega_{\alpha ab} = \Omega_\alpha\varepsilon_{ab}$, where $\Omega_\alpha$ are components of a covector (i.e.\ a 1-form) and $\varepsilon_{ab}$ are components of the Levi-Civita (i.e.\ fully antisymmetric) symbol in two dimensions. Therefore, the spin connection is specified by a single covector. This means that we can write the directional derivative as $\nabla_{\mathbf{v}}\mathbf{n}_i = \dot{u}^\alpha\partial_\alpha \mathbf{n}_i - \dot{u}^\alpha\Omega_\alpha\mathbf{N}_i\times\mathbf{n}_i=\dot{\mathbf{n}}_i - \Omega_t\mathbf{N}_i\times\mathbf{n}_i$ with the scalar $\Omega_t = \dot{u}^\alpha\Omega_\alpha$. The scalar $\Omega_t$ represents the instantaneous angular speed with which the co-moving frame rotates around the local surface normal as $\mathbf{n}_i$ is being transported along the surface. Therefore, in the equation for the dynamics of $\mathbf{n}_i$, we need to subtract the frame's rotation to get the physical change of the vector in the surface's local rest frame. Finally, we note that the covector defining the spin connection is not unique since it is not invariant to rotations of the frame basis around the local surface normal. However, if the local frame basis is rotated by an angle $\theta$, $\Omega_\alpha \to \Omega_\alpha + \partial_\alpha\theta$ (Appendix \ref{app:C}). This is a pure gauge transformation, and the equation of motion for $\mathbf{n}_i$ is invariant under it. The spin connection covector, therefore, plays a role analogous to the vector potential in electromagnetism. One can show that the Gaussian curvature $K=\varepsilon^{\alpha\beta}\partial_\alpha\Omega_\beta$, where $\varepsilon^{\alpha\beta}=\varepsilon_{\alpha\beta}/\sqrt{g}$, and $g$ is the determinant of the metric tensor (Appendix \ref{app:A}) \cite{bowick2009two}.

Using the spin connection, the equation of motion for vector $\mathbf{n}_i$ becomes
\begin{equation}
    \dot{n}_i^a = \left(\boldsymbol{\tau}_i\cdot\mathbf{N}_i +  \Omega_t\right)\left(\mathbf{N}_i\times\mathbf{n}_i\right)^a  
\end{equation}
If we write $\mathbf{n}_i=\cos(\vartheta_i)\mathbf{E}_1+\sin(\vartheta_i)\mathbf{E}_2$, it is straightforward to show that
\begin{equation}
    \dot{\vartheta_i} = \boldsymbol{\tau}_i\cdot\mathbf{N}_i + \Omega_t.
    \label{eq:theta-curved}
\end{equation}
Eqs.\ \eqref{eq:r-curved} and \eqref{eq:theta-curved} fully describe overdamped dynamics of the active agent $i$ confined to move on a curved surface in $\mathbb{R}^3$. It is, thus, evident that the effect of curvature on the $\vartheta_i$ dynamics is to apply a ``geometric torque''.

While the curvature explicitly enters the equation for $\mathbf{n}_i$ via the spin connection, in the overdamped limit, it only appears implicitly in Eq.\ \eqref{eq:r-curved} via operators that project the forces onto the local tangent plane. However, if one is to retain the inertial terms, curvature-related terms would explicitly appear in the expression for the acceleration $\mathbf{a}=\frac{\mathrm{D}\mathbf{v}}{\mathrm{d}t}$.

We note that while using curvilinear coordinates reveals the role of geometry on an agent's trajectory, in numerical simulations, using this description is usually impractical. Namely, for a general surface, curvilinear coordinates cannot be defined globally (e.g.\ poles of a sphere) and one needs to introduce different local charts and supply the appropriate chart transition maps, which can be quite involved. Furthermore, connection coefficients (i.e.\ Christofell symbols and spin connection) are often quite elaborate non-linear functions that are costly to compute at every time step for every agent. Instead, it is usually more practical to handle the surface confinement as a constraint \cite{Sknepnek2015, Henkes2018}. In each time step, one of them makes an unconstrained step, followed by a set of appropriate projections back to the surface and its tangent planes. As long as the time step is sufficiently small that the displacements are small compared to the local curvature radii, it approximates motion on the surface well. This approach was used in simulating the systems presented in Sections \ref{subsec:results-B} and \ref{subsec:results-C}.

\section{Results}
\label{sec:results}

We consider three representative examples specifically chosen to highlight key effects of the geometric torque on active motion.

\subsection{Single active agent}
\label{subsec:single-agent}

To showcase how geometric torque forces a single self-propelled particle to follow a geodesic, we consider an active agent confined to move on the surface of a Gaussian ``bump'', i.e.\ an axisymmetric surface generated by rotating a Gaussian around a vertical axis passing through its maximum. Although a Gaussian bump is not directly motivated by biological systems, it serves as a clean and controllable test geometry for isolating the role of geometric torque. Because the curvature is non-negligible only near the apex, this surface provides a natural setting in which to compare motion in curved and effectively flat regions, thereby clarifying the effects of geometric torque. Furthermore, the axisymmetry of the bump enables straightforward computation of geodesics, simplifying both the analysis and the interpretation of particle trajectories. The surface can be parameterised by two curvilinear coordinates $r$ and $\theta$, which, respectively, measure the distance from the axis of rotation and the angle with an arbitrary axis perpendicular to the axis of rotation. Therefore, points on the surface are given by the vector $\mathbf{r}(r,\theta)=(r\cos\theta,r\sin\theta, h(r))\in\mathbb{R}^3$, where $h(r) = A\exp\left(-\frac{r^2}{2\sigma^2}\right)$ is the height function which, assuming that the axis of rotation is aligned with the $z-$axis, measures the elevation from the $xy$ plane in the ambient Euclidean space $\mathbb{R}^3$. Parameter $A$ sets the maximum height at $r=0$, and $\sigma$ controls the width of the bump. Gaussian curvature of the bump is then $K=\frac{A^2}{\sigma^6} e^{-\frac{r^2}{\sigma ^2}} \left(\sigma ^2-r^2\right)/\left(1+\frac{A^2 r^2 e^{-\frac{r^2}{\sigma ^2}}}{\sigma ^4}\right)^2$, and the angular speed due to geometric torque is $\Omega_t=\dot{\theta}\Omega_\theta=-\frac{\dot{\theta}}{\sqrt{1+(h'(r))^2}}$. Note that the covector $\boldsymbol{\Omega}$ has only the non-zero $\theta$ component, i.e.\ the geometric torque acts to deflect the trajectory in the $\theta$ direction. The expression for $\Omega_t$ can be derived from the expression for the spin connection, following the steps outlined in Appendix \ref{app:C}. In this and the following subsection, we express all distances in units of $\sigma$. Key properties of axisymmetric surfaces are summarised in Appendix \ref{app:D}.

\begin{figure}[t]
\includegraphics[width=1\columnwidth]{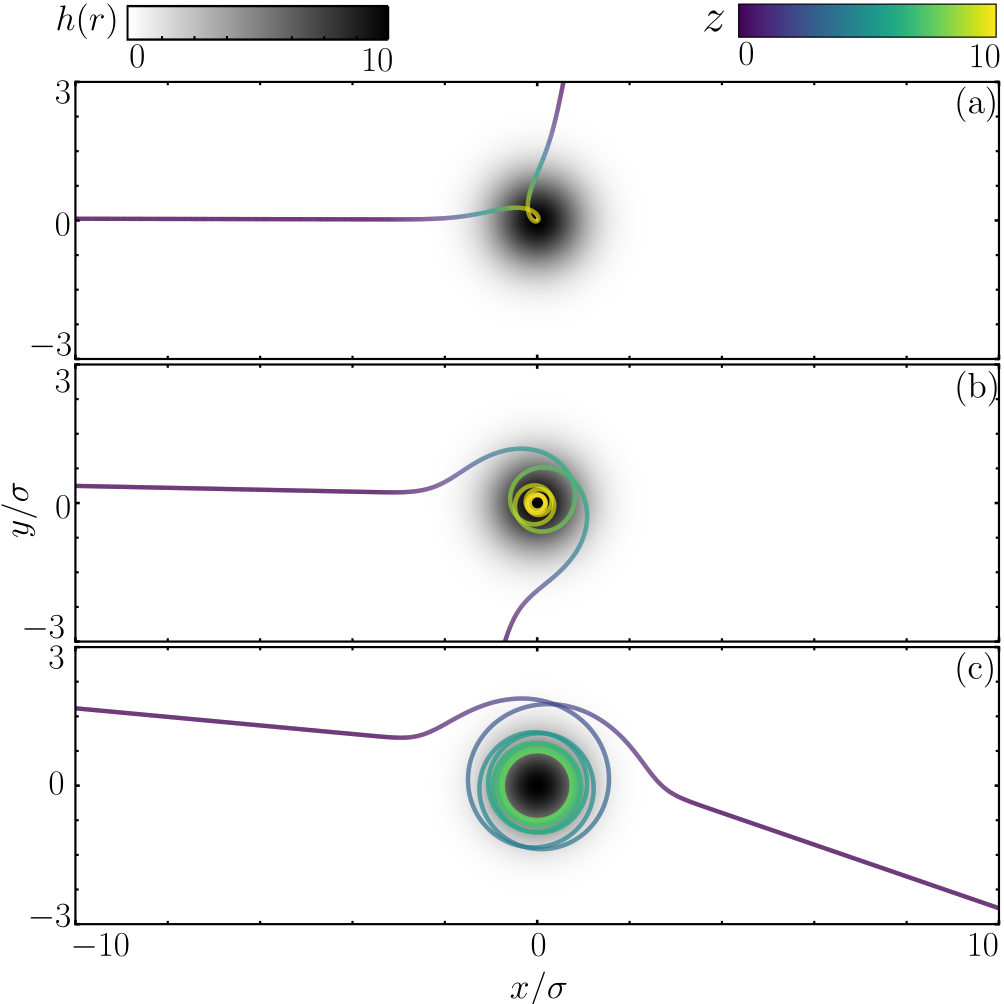}
\caption{\label{fig:lensing_2}Scattering and quasi-trapping of active particles by a Gaussian bump with $A/\sigma=50$. Panel (a) shows the scenario where the Gaussian bump scatters an active particle. Panels (b) and (c) show two cases where an active particle circles around the centre of the bump for an extended time. While there are no closed geodesics on a Gaussian bump, some can wind multiple times, corresponding to the particle being effectively trapped. Shades of grey show the elevation of the Gaussian bump. The parula colours along the trajectory indicate the height which the particle reached.  }
\end{figure}

As we showed in Sec.\ \ref{sec:model}, the active particle follows a geodesic without external torques and noise. Therefore, to understand the motion of the particle, we need to understand geodesics on the surface. A Gaussian bump is convenient since it is not compact (i.e.\ it extends to infinity). Still, the fast exponential decay means the curvature effectively vanishes already at distances exceeding several $\sigma$ from the origin. Therefore, the curvature is localised and can be tuned by tuning the ratio $\sigma/A$. The surface is also geodesically complete \cite{do2016differential}, which, roughly speaking, means that all geodesics are well-behaved. Furthermore, the geodesic equations have two first integrals that have simple physical interpretations, allowing insights into the behaviour of geodesics even without explicitly solving the equations. Although there are no closed geodesics - parallels (i.e.\ circles of constant elevation) are not geodesics, and all geodesics eventually escape to infinity (Appendix \ref{app:D}) - some geodesics wind around the origin multiple times (Fig.\ \ref{fig:lensing_2}). This means the active particle can spend a prolonged time effectively trapped by the bump before escaping.

Next, we analyse the Jacobi field \cite{do2016differential} for several geodesics on a Gaussian bump. For a family of geodesics, $\boldsymbol{\gamma}_\mu(s)$, parametrised by $\mu\in\left(-a,a\right)$ with $a>0$, the Jacobi field measures how the geodesics in the family behave relative to each other. It is defined as $\boldsymbol{\mathcal{J}}(s)=\left.\frac{\partial\boldsymbol{\gamma}_\mu(s)}{\partial \mu}\right|_{\mu=0}$, where $s$ is the arc-length parameter of the geodesic $\boldsymbol{\gamma}_0(s)$ (Appendix \ref{app:E}). In two dimensions, the Jacobi field depends on the Gaussian curvature. It lies in the tangent plane and is orthogonal to the tangent vector, i.e.\ it can be described by a scalar quantity $\mathcal{J}(s)$ such that $\boldsymbol{\mathcal{J}}(s)=\mathcal{J}(s)\mathbf{l}$ where $\mathbf{l}$ is the unit length tangent normal to $\boldsymbol{\gamma}_0(s)$. Therefore, $\mathcal{J}(s)$ measures how the signed distance between nearby geodesics changes as one moves along them. From a physical standpoint, the non-uniform geometric torque acts unevenly on nearby particles, leading to changes in their mutual separation along the surface.

\begin{figure}[t]
\includegraphics[width=1\columnwidth]{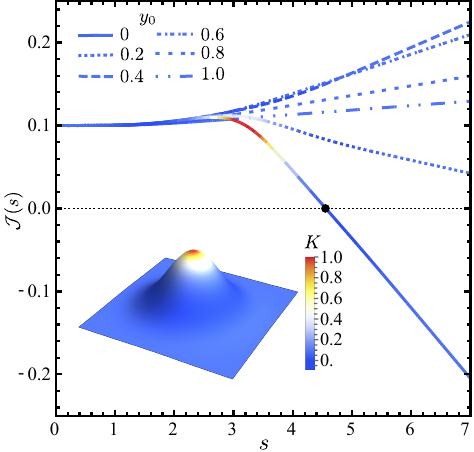}
\caption{\label{fig3} Jacobi field $\mathcal{J}$ as a function of arc-length parameter $s$ for six geodesics. Each geodesic starts at $x_0=-3$ and $y_0\in\{0,0.2,0.4,0.6,0.8,1.0\}$. The Jacobi field is calculated for the ``pair'' geodesic initially at the distance $0.1$ from the corresponding primary geodesic, with both of them initially pointing along $\mathbf{e}_\mathrm{x}$, i.e.\ $\mathcal{J}(0)=0.1$ and $\mathcal{J}'(0)=0$. The Jacobi field curves are coloured by the local value of the Gaussian curvature, $K$. The black dot denotes a conjugate point. $A/\sigma=1$. Inset: Gaussian bump coloured by the local value of Gaussian curvature $K$. Note that for $r<\sigma$ ($r>\sigma$), $K>0$ ($K<0$). The $K<0$ region is, however, exponentially suppressed and has a much weaker effect than the $K>0$ region.}
\end{figure}

In Fig.~\ref{fig3}, we show the Jacobi fields for six pairs of geodesics initially starting far away from the bump, separated by a distance of $0.1$, and all pointing in the $x-$direction with $y(0)\in\{0,0.2,0.4,0.6,0.8,1.0\}$. The Jacobi field strongly depends on the curvature, with the central pair rapidly developing a significant deviation from each other. While the central (i.e.\ $y(0)=0$) geodesic passes straight through the bump, its pair first gets slightly deflected, to be lensed back. It eventually crosses the central geodesic ($\mathcal{J}=0$ -- a conjugate point denoted with the black dot in Fig.\ \ref{fig3}) and continues to diverge. As one moves to higher values of $y(0)$, i.e.\ away from the bump, the deflection is less pronounced with pairs remaining nearby each other. 

\subsection{Flock on a Gaussian bump}
\label{subsec:results-B}
In the previous subsection, we showed that the curvature affects the separation of two nearby geodesics. To illustrate how geometric torque and alignment interactions shape flocking behavior, we consider a simplified scenario: a column of perfectly aligned agents advancing toward a Gaussian bump. Since the model with alignment interactions is not analytically tractable, we resort to numerical simulations based on the projection method, outlined in the introduction and discussed in Ref.\ \cite{Sknepnek2015}.

\begin{figure}[t]
\includegraphics[width=1\columnwidth]{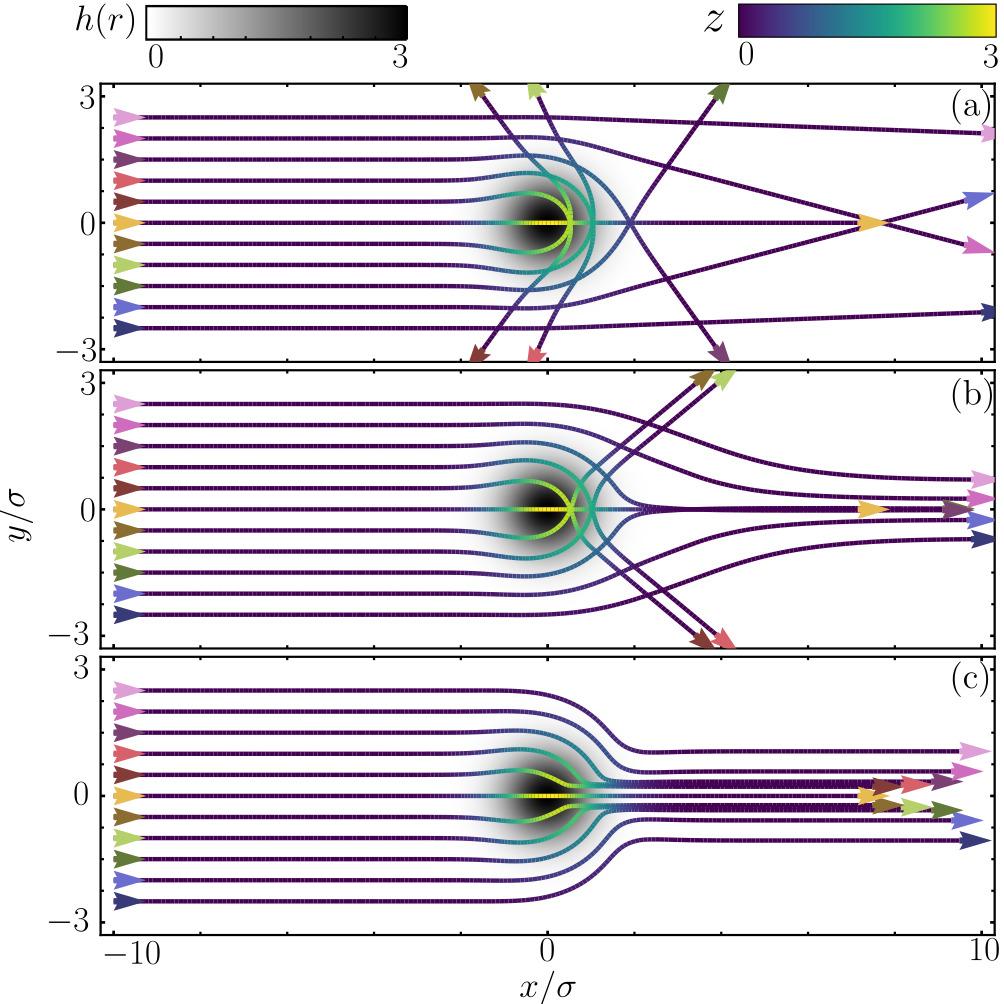}
\caption{\label{fig4} A flock of active agents moving along the Gaussian bump with $A/\sigma=3$. (a) $J=0$ (i.e.\ a non-interacting flock) where particles move purely under the influence of geometric torque and follow geodesics. (b) Intermediate alignment ($J=1.0$) --  particles no longer follow geodesics, and the flock scatters as it crosses the bump. (c) Strong alignment ($J=10.0$), where external torque dominates, resulting in a significantly more compact flock. Coloured triangles represent individual particles before ($x/\sigma = -10$) and after ($x/\sigma = 10$) crossing the bump. The flock is initially composed of $11$ particles, each separated by a distance of $0.5$. Shades of grey represent the height profile of the Gaussian bump. Particle trajectories are coloured according to the ``elevation'' they reach.}
\end{figure}

We start by considering a flock of $N=11$ non-interacting particles, initially placed on a straight line along the $y-$axis with their directors all pointing along the $x-$axis. The particles are sufficiently far from the bump, and the curvature can be neglected. Therefore, at the initial placement of the flock, we do not need to distinguish between the surface and the $xy$ plane of the ambient space. In Fig.\ \ref{fig4}a, we show trajectories for $N=11$ agents colour-coded for easier tracking. In the absence of alignment interactions, particle motion is solely governed by the geometric torque, i.e.\ each particle follows a geodesic. Geodesics close to the bump quickly deviate from each other (Fig.\ \ref{fig3}), and the flock completely disintegrates as it passes over the bump. Notably, trajectories of light blue, dark pink, and lemon-coloured particles intersect at the conjugate point (black dot in Fig.\ \ref{fig3}). This behaviour further illustrates why the Gaussian bump is a convenient test geometry. As particles move away from the bump centre, the curvature and, therefore, the geometric torque decay rapidly, so geodesics asymptotically approach straight lines. This follows from the expression $\Omega_t=\dot{\theta}/\sqrt{1-\frac{4A^2 r^2}{\sigma^4}\exp\left(-\frac{2r^2}{\sigma^2}\right)}$ whose denominator approaches unity exponentially fast as $\frac{r}{\sigma} \rightarrow \infty$. Consequently, the influence of geometric torque, and hence the deviation from straight motion, vanishes quickly outside the central curved region.

We then introduce an alignment torque between agents within a cutoff distance $r_\mathrm{cut}=1$, calculated in the embedding $\mathbb{R}^3$ space. The torque on agent $i$ is $\boldsymbol{\tau}_i=\frac{J}{N_i}\sum_{j\in\mathcal{N}_i}\mathbf{n}_i\times\mathbf{n}_j$, where $J$ is the alignment strength (i.e.\ $J^{-1}$ sets the alignment timescale), and $N_i$ is the number of agents in the $r_\mathrm{cut}-$neighbourhood $\mathcal{N}_i$ of particle $i$. The cross product is also calculated in $\mathbb{R}^3$, and the resulting torque is projected on the local surface normal $\mathbf{N}_i$. To avoid complications due to excluded volume effect, all steric interactions are neglected. In Fig.\ \ref{fig4}b, we show the trajectories for an intermediate value of $J=1$. The geometric torque competes with the alignment, altering the trajectories. Particles no longer follow geodesics, but the curvature is strong enough to scatter them, effectively dispersing the flock. Finally, at $J=10$, the alignment torque dominates, preventing the flock from falling apart (Fig.\ \ref{fig4}c). The trajectories converge, but some agents lag behind the rest of the flock. This is due to particles needing to travel different distances over vs.\ around the bump.

Therefore, even in the presence of strong alignment, curvature acts to disrupt the ordered, flock-like collective motion. This, however, does not preclude the possibility of devising a geometric pattern that would allow focusing or even sorting of agents based on, e.g.\ their speed. 

\subsection{Band-state on a sphere}
\label{subsec:results-C}

In this section, we revisit an interesting aspect of the dynamics of a collective of self-propelled agents confined to move on the surface of a 2-sphere \cite{Sknepnek2015}. In this model, agents are placed on a sphere, propelled along unit vectors $\mathbf{n}_i$, and interact through a short-range soft repulsive potential; with nearest-neighbor alignment, they collectively form a rotating band state \cite{Sknepnek2015}. Interestingly, the orientation vectors $\mathbf{n}_i$ form an angle $\alpha$ with the lines of latitude that varies linearly with the latitude angle $\psi$ near the equator, but deviates from linearity close to the band edges \cite{Sknepnek2015}. Equivalently, $\mathbf{n}_i$ tilt progressively towards the band centre away from the equator. According to Ref.\ \cite{Sknepnek2015}, the inward push from the band edges generates the equatorial density maximum. To capture this effect, they introduced a simplified ``rotating slice'' model, approximating the band by a linked-spring chain along a meridian that rotates uniformly around the polar axis. The model is then solved numerically with the $\alpha$ vs.\ $\phi$ dependence not explained but obtained by fitting simulation data. Here, we show how the $\alpha$ vs.\ $\phi$ curve for the ``rotating slice'' model can be obtained using the geometric torque.

\begin{figure}[t]
\includegraphics[width=\columnwidth]{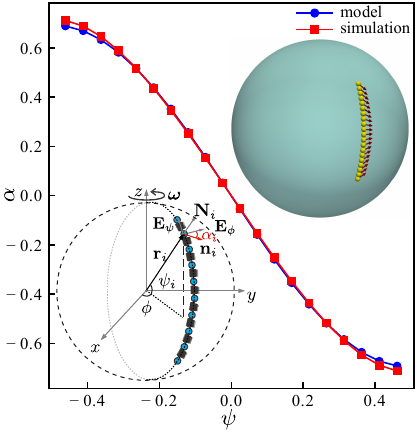}
\caption{\label{fig5} Angle $\alpha$ between $\mathbf{n}_i$ and the $\mathbf{E}_\phi$ axis of the local orthonormal frame vs.\ the latitude $\psi$ for the numerical solution (blue) and the simulation (red) of the toy model for interacting active agents on a sphere of radius $R=10$, with $J=0.5$ and $v_0=0.5$, $k=100$, $\sigma=0.25$. Bottom inset: Sketch of the toy model of the band. Blue discs represent particles and grey coils represent the soft-core potential between nearest neighbours. Top inset: Snapshot of the simulation showing positions of agents (yellow) and directions of the vectors $\mathbf{n}_i$ (red arrows).}
\end{figure}

More precisely, the one-dimensional ``rotating slice'' model assumes a slice through the band along a meridian. The slice rotates around the $z-$axis (assumed to pass through the north pole) at a constant angular velocity $\omega$. All particles along the slice can move up and down the meridian but cannot advance or lag relative to other particles (i.e.\ mixing within the band is not possible). Neighbouring particles interact via the soft potential with stiffness $k$ and rest length $2\sigma_\mathrm{d}$. Vectors $\mathbf{n}_i$ align to their immediate neighbours with the alignment strength $J$ (Fig.\ \ref{fig5} - lower inset).

Since all particles reside on the same meridian, they all have the same fixed value of coordinate $\phi$. The projection of the torque $\boldsymbol{\tau}_i$ on to the local surface normal $\mathbf{N}_i$ is $\boldsymbol{\tau}_i\cdot\mathbf{N}_i=J\left(\mathbf{n}(\psi_i)\times\mathbf{n}(\psi_{i-1})+\mathbf{n}(\psi_i)\times\mathbf{n}(\psi_{i+1})\right)\cdot\mathbf{N}_i$. Assuming that $\mathbf{n}(\psi_i)=\cos\alpha_i\mathbf{E}_\phi+\sin\alpha_i\mathbf{E}_\psi$, a direct calculation shows that $\boldsymbol{\tau}_i\cdot\mathbf{N}_i=J\left(\sin(\alpha_{i+1}-\alpha_i) - \sin(\alpha_i - \alpha_{i-1})\right)\approx J(\alpha_{i+1} - 2\alpha_i + \alpha_{i-1})$, for $\left|\alpha_i-\alpha_{i\pm1}\right|\ll1$. The spin connection covector has only one non-zero component, $\Omega_\phi = -\sin\psi\mathrm{d}\phi$. If we assume that the velocity points in the $\phi-$direction, we have $\mathbf{v}=v^\phi\mathbf{e}_\phi$ (no summation). Note that the component $v^\phi$ does not have the dimensions of $\mathrm{length}\times\mathrm{time}^{-1}$ since $\mathbf{e}_\phi$ is not normalised but contains factors of $R$. Since $\Omega_t=v^\mu\Omega_\mu=-\sin\psi\mathrm{d}\phi(\mathbf{v})=-v^\phi\sin\psi$ (here $\mathrm{d}\phi(\mathbf{v})$ denotes the action of the 1-form $\mathrm{d}\phi$ on vector $\mathbf{v}$), we have the equation of motion for $\alpha_i$ as
\begin{equation}
    \dot{\alpha}_i = J(\alpha_{i+1} - 2\alpha_i + \alpha_{i-1}) - v_i^\phi\sin\psi_i.
\end{equation}
Assuming the rotating band is a steady state, we have $\dot{\alpha_i}=0$, i.e. the alignment torque is balanced by the geometric torque. Therefore,
\begin{equation}
    \alpha_{i+1} - 2\alpha_i + \alpha_{i-1} = \frac{v_i^\phi}{J}\sin\psi_i,
\end{equation}
with the boundary terms $\alpha_2-\alpha_1=\frac{v_1^\phi}{J}\sin\psi_1$ and $-\alpha_N + \alpha_{N-1}=\frac{v_N^\phi}{J}\sin\psi_N$, and with index $i\in\{1,\dots,N\}$. For the position, we have the force balance $\gamma_\mathrm{t}\mathbf{v}_i=f_0\mathbf{n}_i + \mathbf{F}_i$. Assuming that $\left|\psi_i - \psi_{i\pm1}\right|\ll1$ for all $i$, it is straightforward to show that the $\psi-$component of the force is $F_i^\psi=kR\left(\psi_{i+1} - 2\psi_i + \psi_{i-1}\right)$ with boundary terms $F_1^\psi=kR\left(\psi_2-\psi_{1}\right) - 2\sigma_\mathrm{d} k$ and $F_N^\psi=-kR\left(\psi_N-\psi_{N-1}\right) + 2\sigma_\mathrm{d} k$. The force balance in the $\phi-$direction leads to $v_i^\phi=\frac{v_0}{R}\frac{\cos\alpha_i}{\cos\psi_i}$, where $v_0=f_0/\gamma_\mathrm{t}$. Combining all results leads to a closed system of nonlinear equations for the steady state values of $\alpha_i$ and $\psi_i$,
\begin{subequations}
    \begin{eqnarray}
        \alpha_{i+1} - 2\alpha_i + \alpha_{i-1}  &=& \frac{v_0}{JR}\cos\alpha_i\tan\psi_i,\label{eq:1d-band-a}\\
        \psi_{i+1} - 2\psi_i + \psi_{i-1} &=& -\frac{\gamma_\mathrm{t}v_0}{kR}\sin\alpha_i, \label{eq:1d-band-b}
    \end{eqnarray}
\end{subequations}
with boundary terms $\alpha_2-\alpha_1=\frac{v_0}{JR}\cos\alpha_1\tan\psi_1$, $-\alpha_N + \alpha_{N-1}=\frac{v_0}{JR}\cos\alpha_N\tan\psi_N$, $\psi_2-\psi_1=-\frac{\gamma_\mathrm{t}v_0}{kR}\sin\alpha_1 + \frac{2\sigma_\mathrm{d}}{R}$, and $-\psi_N+\psi_{N-1}=-\frac{\gamma_\mathrm{t}v_0}{kR}\sin\alpha_N - \frac{2\sigma_\mathrm{d}}{R}$. For a given set of parameters $v_0$ and $J$ (assuming that $k$, $R$, and $\gamma_\mathrm{t}$ are fixed), these equations can be solved numerically. Therefore, using this picture of geometric torque, we are able to model the physics of the rotating band state without resorting to the empirically observed relationship between $\alpha$ and $\psi$.

To test our semi-analytical predictions, we simulated a chain of $N=20$ beads interacting with immediate neighbours via a harmonic potential $V\left(r_{i,i\pm1}\right)=\frac{k}{2}\left(r_{i,i\pm1}-2\sigma_\mathrm{d}\right)^2$ with $k=100$ and $\sigma_{\mathrm{d}} = 0.25$. The chain is placed on a sphere of radius $R=10$ (Fig.\ \ref{fig5} - top inset). Each bead has a unit-length vector $\mathbf{n}_i$ tangent to the sphere at every point and is propelled in the direction $\mathbf{n}_i$ with speed $v_0=0.5$. $\mathbf{n}_i$ vectors of neighbouring points are subject to an alignment torque of magnitude $J=0.5$, without noise. Finally, we introduced an angle potential of strength $k_{\mathrm{ang}} = 5$ between each consecutive triplet of beads to ensure all beads remained on the same meridian. The overdamped equations of motion were integrated using the forward Euler method with time step $\delta t = 0.001$. 

In Fig.\ \ref{fig5}, we show the numerical solution (blue curve) of Eqs.\ \eqref{eq:1d-band-a} and \eqref{eq:1d-band-b} and the simulations (red curve) of the chain model discussed in the previous paragraph. The simulations and numerical solution of Eqs.\ \eqref{eq:1d-band-a} and \eqref{eq:1d-band-b} show an excellent agreement with each other. This confirms that Eqs.\ \eqref{eq:1d-band-a} and \eqref{eq:1d-band-b} correctly capture the physics of the ``rotating-slice'' model, showing that the observed profile can be understood as an interplay of the geometric torque, alignment and steric repulsion.  

We remark that while the ``rotating-slice'' model can qualitatively capture the $\alpha$ vs.\ $\phi$ dependence observed in full-scale simulations of self-propelled agents on a sphere, it fails to do so qualitatively. Analysis of trajectories of individual agents in the full simulation shows that particles do not travel on circles coinciding with lines of constant latitude, as assumed by the simplified model, but follow more complex loops that traverse a significant portion of the band's width. The motion is not turbulent, but leads to mixing within the band, which can be confirmed by tracking the relative distance of pairs of particles over time. The band state is, therefore, liquid-like. A detailed analysis of this state is, however, beyond the scope of this work and will be addressed elsewhere.

\section{Discussion and conclusions}
\label{sec:discussion}

In this study, we employed a model of self-propelled particles -- a class of active matter systems -- to investigate how curvature influences both single-particle and collective dynamics. In the absence of external forces, torques, or noise, a particle moves along a geodesic, the curved space analogue of a straight line. This effect can be captured by the notion of a geometric torque, which reflects the intrinsic coupling between translations and rotations on a curved surface, as also discussed in Ref.\ \cite{nemeth2025intrinsic}. Unlike straight lines in flat space, which are either parallel or intersect once, geodesics can exhibit a much richer range of behaviours. Depending on the geometry and topology of the surface, they may self-intersect, form closed loops (e.g.\ as in great circles on a sphere), or diverge. Consequently, the motion of a single active particle in curved space becomes significantly more complex, giving rise to phenomena such as curvature-induced lensing and trapping.

Curvature also affects flocking. A flock of perfectly aligned agents can get dispersed by curvature since the distance between neighbouring geodesics is not constant but depends on the local Gaussian curvature. This effect is analogous to the tidal forces on a group of free-falling non-interacting particles in the gravitational field. The convergence or divergence of geodesics also leads to misalignment in the direction of self-propulsion. Even in the presence of strong alignment interactions, the flock loses coherence, as neighbouring particles must traverse different distances when moving across a curved region. For instance, on a Gaussian bump, it takes longer to travel over the peak than to pass along its sides, resulting in spatially varying speeds and disrupted coordination. Therefore, particles that started shoulder to shoulder would separate despite moving at the same speed $v_0$ and having perfectly aligned directions in $\mathbb{R}^3$.

In many experimental systems, active agents are not point-like particles, but often rather stiff filaments. The role of activity in the behaviour of filaments is, however, only partly understood even in the flat case \cite{isele2015self, prathyusha2018dynamically,duman2018collective,winkler2020physics,fazelzadeh2023effects,janzen2024active}. Even without activity, a semi-flexible filament confined to a curved surface, however, does not necessarily align with a geodesic \cite{nickerson1988intrinsic}. While the proof is elaborate \cite{nickerson1988intrinsic,guven2014environmental}, a physical argument involves comparing geodesic and normal curvatures. Namely, the local curvature of the filament can be split into normal and tangential components along the confining surface. In many situations, bending in the tangential direction (i.e.\ deviating from a geodesic) is favourable to minimise bending in the normal direction. This makes the interaction between filament and surface curvatures intricate. It is completely unknown how such effects couple to activity.

The sources of activity, however, go beyond self-propulsion, e.g.\ growth, cell division and death, cell and tissue-level active stresses due to the action of molecular motors on cytoskeletal filaments, etc. These are ubiquitous in biological systems, especially during embryonic development. For example, the establishment of the body plan relies on cells taking the correct position, which requires large-scale collective migration, driven by activity. In this case, curvature develops concomitantly with cell migration, making it an inseparable part of the mechanics. Understanding how curvature and activity coordinate to form complex functional shapes is only partly understood.

%------------- Acknowledgments -------------%
\begin{acknowledgments}
We thank Demian Levis for helpful discussions during the early stages of this project. E.D.M.\ acknowledges funding by the endowed E.N.\&M.N.\ Lindsay PhD studentship. R.S.\ acknowledges support from the UK Engineering and Physical Sciences Research Council (Award EP/W023946/1). DMF and GJ acknowledge support from the Comunidad de Madrid and the Complutense University of Madrid (Spain) through the Atracción de Talento program (Grant No. 2022-T1/TIC-24007). DMF also acknowledges support from MINECO (Grant No. PID2023-148991NA-I00).
\end{acknowledgments}

\appendix

\section{A brief overview of the geometry of 2D surfaces embedded in $\mathbb{R}^3$}
\label{app:A}

Here we provide a brief overview of the theory of surfaces embedded in the Euclidean space $\mathbb{R}^3$. A detailed account can be found in standard textbooks, e.g.\ \cite{do2016differential,spivak1970comprehensive}

A smooth two-dimensional surface embedded in $\mathbb{R}^3$ is described by a function $\mathbf{r}:D\to\mathbb{R}^3$, with $D\subset\mathbb{R}^2$, i.e.\ for the pair of curvilinear coordinates $(u,v)\in D$, $(u,v)\mapsto\mathbf{r}(u,v)$. If one chooses Cartesian coordinates in $\mathbb{R}^3$, $\mathbf{r}(u,v) = (x(u,v), y(u,v), z(u,v))$. For example, $\mathbf{r}(u,v)=(u\cos v,u \sin v,v)$, with $(u,v)\in\mathbb{R}^2$ describes a helicoid. It is also assumed that $\mathbf{r}(u,v)$ is differentiable a sufficient number of times at any point in $D$ (smoothness) and that at every point the vectors $\frac{\partial\mathbf{r}}{\partial u}$ and $\frac{\partial\mathbf{r}}{\partial v}$ are linearly independent (non-degeneracy). In general, a single parametrisation $\mathbf{r}(u,v)$ might not cover the entire surface without encountering coordinate singularities, especially if the surface is topologically nontrivial. A typical example is a sphere parameterised by the latitude ($u\equiv\theta$) and longitude ($v\equiv\varphi$), where $\varphi$ is ill-defined at the poles, i.e.\ for $\theta\in\{0,\pi\}$. In such cases, one requires multiple overlapping coordinate charts. 

One can define two tangent vectors $\mathbf{e}_u=\frac{\partial\mathbf{r}}{\partial u}$ and $\mathbf{e}_v=\frac{\partial\mathbf{r}}{\partial v}$ at every point of the surface, which span local tangent planes. Vectors $\mathbf{e}_u$ and $\mathbf{e}_v$ can be understood as living in $\mathbb{R}^3$ but constrained to be tangent to the surface (extrinsic view) or as being a basis of the local two-dimensional tangent space (intrinsic view) with no reference to the embedding space. In general, $\mathbf{e}_u$ and $\mathbf{e}_v$ are neither orthogonal to each other nor unit length. If we introduce $u\equiv u^1$ and $v\equiv u^2$, we can write $\mathbf{e}_\alpha=\frac{\partial\mathbf{r}}{\partial u^\alpha}\equiv\partial_\alpha\mathbf{r}$ with $\alpha\in\{1,2\}$. A vector $\mathbf{a}$ in the tangent plane at a point $P$ can be written as $\mathbf{a}=a^\alpha\mathbf{e}_\alpha$, where $a^\alpha$ are coordinates of $\mathbf{a}$ in the coordinate tangent basis $\{\mathbf{e}_{u^1},\mathbf{e}_{u^2}\}$, and we assume summation over pairs of repeated upper and lower indices. We use upper Greek indices to denote coordinates of a tangent vector in the coordinate tangent basis. 

The induced metric (or the first fundamental form) on the surface is a type-$(0,2)$ tensor with components
\begin{equation}
    g_{\alpha\beta} = \mathbf{e}_\alpha\cdot\mathbf{e}_\beta,
\end{equation}
where $\cdot$ is the dot product in $\mathbb{R}^3$. It allows one to compute angles in the tangent plane and distances on the surface. One can also define the inverse metric tensor with components $g^{\alpha\beta}$. This $(2,0)$ tensor is defined as $g^{\alpha\gamma}g_{\gamma\beta}=\delta^{\alpha}_\beta$, where $\delta^{\alpha}_\beta$ is the Kronecker's delta. The metric and the inverse metric tensors can be used to ``lower'' and ``raise'' indices, i.e.\ $b_\alpha = b^\beta g_{\alpha\beta}$ and $b^\alpha = b_\beta g^{\alpha\beta}$. Here, $b_\beta$ are coefficients of a 1-form (i.e.\ a covector) expressed in the coordinate tangent basis. Using these expressions, the dot product between two tangent vectors $\mathbf{a}=a^\alpha\mathbf{e}_\alpha$ and $\mathbf{b}=b^\beta\mathbf{e}_\beta$ at a given point is $\mathbf{a}\cdot\mathbf{b}=a^\alpha b^\beta\mathbf{e}_\alpha\cdot\mathbf{e}_\beta=g_{\alpha\beta}a^\alpha b^\beta=a^\alpha b_\alpha$. 

Since we'll refer to both the surface and the embedding $\mathbb{R}^3$ space, we will use the extrinsic view. The normal vector to the surface at the point $P$ is defined as 
\begin{equation}
   \label{eq:surface_normal}
   \mathbf{N}=\frac{\mathbf{e}_u\times\mathbf{e}_v}{\left|\mathbf{e}_u\times\mathbf{e}_v\right|}, 
\end{equation} 
where $\left|\cdot\right|$ is the Euclidean norm. The direction of $\mathbf{N}$ defines the orientation of the surface. We restrict attention to orientable surfaces, where a consistent choice of normal is possible globally. One can then define the second fundamental form, a $(0,2)$ tensor with components 
\begin{equation}
    b_{\alpha\beta} = -\mathbf{e}_\alpha\cdot\partial_\beta\mathbf{N}.
\end{equation}
The $(1,1)$ curvature tensor has components $C^\alpha_\beta=g^{\alpha\gamma}b_{\beta\gamma}$. It is symmetric, hence we do not need to distinguish the order in which indices appear. Half the trace of the curvature tensor is the mean curvature, $H$, and its determinant is the Gaussian curvature, $K$. These two quantities play an important role in modelling membranes \cite{nelson2004statistical}. The formula of Weingarten relates the curvature to the spatial derivatives of the surface normal as $\partial_\alpha\mathbf{N} = -C^\beta_\alpha\mathbf{e}_\beta$, reflecting the intuitive notion that the direction of the normal vector changes as one moves along a curved surface.

It is often required to compute derivatives of tangent vectors. In general, for a tangent vector field $\mathbf{a}$, the derivative $\partial_\alpha\mathbf{a}$ is a vector that is not necessarily tangent to the surface. Instead, we have $\partial_\alpha\mathbf{a} = \partial_\alpha(a^\beta\mathbf{e}_\beta)=(\partial_\alpha a^\beta)\mathbf{e}_\beta + \tensor{\Gamma}{^\beta_{\alpha\gamma}}a^\gamma\mathbf{e}_\beta + a^\beta b_{\alpha\beta}\mathbf{N}$. The part of vector $\partial_\alpha\mathbf{a}$ that remains in the tangent plane is called the covariant derivative and is denoted as 
\begin{equation}
 \nabla_\alpha\mathbf{a}=\left(\partial_\alpha a^\beta + \tensor{\Gamma}{^\beta_{\alpha\gamma}}a^\gamma\right)\mathbf{e}_\beta.   
\end{equation}
It plays a central role in Riemannian geometry. $\tensor{\Gamma}{^\alpha_{\beta\gamma}}$ is the Christoffel symbol of the second kind, satisfying $\nabla_\alpha\mathbf{e}_\beta=\tensor{\Gamma}{^\gamma_{\alpha\beta}}\mathbf{e}_\gamma$. Christoffel symbols, therefore, describe how the coordinate tangent basis changes as one moves along the surface, i.e.\ they specify how different tangent planes are connected. Therefore, they describe the Levi-Civita connection \cite{lee2018introduction} on the surface expressed in the coordinate basis. Hence, they are also known as connection coefficients. The Christoffel symbols $\tensor{\Gamma}{^\alpha_{\beta\gamma}}$ can be expressed in terms of derivatives of the metric tensor as $\tensor{\Gamma}{^\alpha_{\beta\gamma}} = \frac{1}{2} g^{\alpha\delta} \left( \partial_\gamma g_{\delta\beta} + \partial_\beta g_{\delta\gamma} - \partial_\delta g_{\beta\gamma} \right)$. They, however, are not components of a rank-3 tensor. 

The covariant derivative of a vector field $\mathbf{a}$ in the direction of a tangent vector $\mathbf{b}=b^\alpha\mathbf{e}_\alpha$ is given as $\nabla_{\mathbf{b}}\mathbf{a}=b^\alpha\nabla_\alpha\mathbf{a}=\left(b^\alpha\partial_\alpha  a^\beta + \tensor{\Gamma}{^\beta_{\alpha\gamma}}b^\alpha a^\gamma\right)\mathbf{e}_\beta$. If $\nabla_\mathbf{b}\mathbf{a}=0$, vector $\mathbf{a}$ is parallel transported along vector $\mathbf{b}$. On a curved surface, the idea of being ``parallel'' depends on the path of transport. Namely, two vectors can remain parallel when moved along one curve yet fail to stay parallel when carried along a different curve. 
\section{Curves on a surface and the geodesic equation}
\label{app:B}

A curve embedded in a surface is described by a vector $\boldsymbol{\gamma}(s)=\mathbf{r}(u(s),v(s))$ parametrised by a single parameter $s$, where $s$ can be, e.g.\ arc-length parameter, time, or some other suitable quantity. The tangent vector to the curve is $\mathbf{t}=\frac{\mathrm{d}\boldsymbol{\gamma}}{\mathrm{d}s}$. Using the chain rule, $\mathbf{t}=\frac{\mathrm{d}\boldsymbol{\gamma}}{\mathrm{d}s}=\frac{\mathrm{d}u^\alpha}{\mathrm{d}s}\partial_\alpha\mathbf{r}$, where $\dot{\gamma}^\alpha\equiv\frac{\mathrm{d}u^\alpha}{\mathrm{d}s}$. Therefore, vector $\mathbf{t}$ is tangent to the surface. If the parameter $s$ is time, the functions $\dot{\gamma}^\alpha(s)$ are components of the velocity vector along the curve. We only consider regular curves, i.e. $\left|\mathbf{t}\right|>0$ at every point. In general, $\left|\mathbf{t}\right|\neq1$. If, however, $\left|\mathbf{t}\right|=1$ at every point, $s$ is called the arc-length parameterization. Many formulas are simplified in the arc-length parametrisation. 

In the arc-length parametrisation, the curvature of a curve is a vector $\mathbf{k}=\frac{\mathrm{d}\mathbf{t}}{\mathrm{d}s}$. It can be split into the normal and tangential components as $\mathbf{k}=k_g\mathbf{l} + k_n\mathbf{N}$, where $\mathbf{l}=\mathbf{N}\times\mathbf{t}$ is the tangent normal, i.e. a vector tangent to the surface and orthogonal both to $\mathbf{t}$ and $\mathbf{N}$. $k_n$ is the scalar normal curvature, and it measures how much the curve is curved due to the curvature of the embedding surface. $k_g$ is the scalar geodesic curvature,  and it measures the intrinsic curvature of the curve. In a plane, $k_n=0$ and $k_g$ measures how much the curve deviates from a straight line. Since $\mathbf{l}$ and $\mathbf{N}$ are orthonormal, $\left|\mathbf{k}\right|^2 = k_n^2 + k_g^2$.

Let $\mathbf{a}(s)$ be a vector field defined along the curve $\boldsymbol{\gamma}$. The derivative along the curve is $\frac{\mathrm{D}\mathbf{a}}{\mathrm{d}s}\equiv \mathrm{D}_s\mathbf{a}=\nabla_{\mathbf{t}}\mathbf{a}=\dot{\gamma}^\alpha\nabla_\alpha\mathbf{a}=\left(\dot{\gamma}^\alpha\partial_\alpha a^\beta + \tensor{\Gamma}{^\beta_{\alpha\gamma}}\dot{\gamma}^\alpha a^\gamma\right)\mathbf{e}_\beta$. One can show that the geodesic curvature of curve $\boldsymbol{\gamma}$ is $\mathbf{k}_g\equiv k_g\mathbf{l}=\nabla_{\mathbf{t}}\mathbf{t}\equiv\nabla_{\dot{\boldsymbol{\gamma}}}\dot{\boldsymbol{\gamma}}$. The case $\nabla_{\mathbf{t}}\mathbf{t}=0$, defines a geodesic, i.e.\ a curve along which the tangent vector is parallel transported. In the coordinate tangent basis, geodesics are solutions of a set of second-order (typically nonlinear) ordinary differential equations
\begin{equation}
    \frac{\mathrm{d}^2\gamma^\alpha}{\mathrm{d}s^2}+\tensor{\Gamma}{^\alpha_{\beta\gamma}}\frac{\mathrm{d}\gamma^\beta}{\mathrm{d}s}\frac{\mathrm{d}\gamma^\gamma}{\mathrm{d}s}=0.
\end{equation}
Specifying $\gamma^\alpha(0)$ and $\dot{\gamma}^\alpha(0)$ determines geodesic uniquely. Loosely speaking, if two points on the surface are sufficiently close, the geodesic connecting them minimises the distance measured along the surface. Therefore, on the surface, geodesics play the role of a straight line in a flat space. Unlike straight lines, however, geodesics can self-intersect multiple times or even be closed loops (e.g.\ the equator of a sphere). 

\section{Tetrads and spin connection}
\label{app:C}
In Appendix \ref{app:A}, we introduced the coordinate tangent basis as $\mathbf{e}_\alpha = \partial_\alpha\mathbf{r}$. In general, the coordinate basis is not orthonormal. One can, e.g.\ using the Gram-Schmidt procedure, introduce an orthonormal basis $\mathbf{E}_a$, with $a\in\{1,2\}$. This basis is sometimes known as the local frame. We'll use Latin indices from the beginning of the alphabet (e.g.\ $a$, $b$, $c$,...) to denote components in this basis, such that a tangent vector $\mathbf{q}=q^\alpha\mathbf{e}_\alpha=q^a\mathbf{E}_a$. The two bases are connected by a basis change matrix $\tensor{\Lambda}{^\alpha_a}$ such that $\mathbf{E}_a=\tensor{\Lambda}{^\alpha_a}\mathbf{e}_\alpha$, with the inverse transform $\mathbf{e}_\alpha=\tensor{\Lambda}{^a_\alpha}\mathbf{E}_a$. $\tensor{\Lambda}{^a_\alpha}$ are components of the zweibein, and $\tensor{\Lambda}{^a_\alpha}\tensor{\Lambda}{^\alpha_b}=\delta^a_b$ and $\tensor{\Lambda}{^\alpha_a}\tensor{\Lambda}{^a_\beta}=\delta^\alpha_\beta$. Since $\mathbf{E}_a\cdot\mathbf{E}_b=\delta_{ab}$, where $\delta_{ab}$ is Kronecker's delta, in the orthonormal basis, the metric is flat. One can use zweibeins to relate components of the metric tensor in the orthonormal basis to components of the metric tensor in the coordinate basis as $g_{\alpha\beta}=\delta_{ab}\tensor{\Lambda}{^a_\alpha}\tensor{\Lambda}{^b_\beta}$. A more detailed discussion can be found, e.g.\ in Refs.\ \cite{kamien2002geometry, bowick2009two, wald2010general}.

The covariant derivative of a vector $\mathbf{q}=q^a\mathbf{E}_a$ written in the orthonormal basis is
\begin{equation}
    \nabla_\alpha\mathbf{q} = \left(\partial_\alpha q^a + \tensor{\omega}{_\alpha^a_b}q^b\right)\mathbf{E}_a,
\end{equation}
where $\tensor{\omega}{_\alpha^a_b}$ are coefficients of the so-called spin connection. Analogous to the Christoffel symbols, $\nabla_\alpha\mathbf{E}_b=\tensor{\omega}{_\alpha^a_b}\mathbf{E}_a$, which gives $\mathbf{E}_a\cdot\nabla_\alpha\mathbf{E}_b=\tensor{\omega}{_\alpha^c_b}\mathbf{E}_a\cdot\mathbf{E}_c=\tensor{\omega}{_\alpha^c_b}\delta_{ac}=\omega_{\alpha ab}$. Applying the product rule for covariant derivatives to $\nabla_\alpha\left(\mathbf{E}_a\cdot\mathbf{E}_b\right)=0$ immediately shows that the spin connection is antisymmetric in Latin indices, i.e.\ $\omega_{\alpha ab}=-\omega_{\alpha ba}$. In two dimensions, this means that $\omega_{\alpha ab}$ can we written as a product of the antisymmetric symbol $\varepsilon_{ab}=\delta_{a1}\delta_{b2}-\delta_{a2}\delta_{b1}$ and a covector $\Omega_\alpha$, i.e.\ $\omega_{\alpha ab}=\varepsilon_{ab}\Omega_\alpha$. A lengthy but straightforward calculation relates the coefficients of the spin connection to the Christoffel symbols as $\tensor{\omega}{_\alpha^ a_b}=\tensor{\Gamma}{^\gamma_{\beta\alpha}}\tensor{\Lambda}{^a_\gamma}\tensor{\Lambda}{^\beta_b}-\tensor{\Lambda}{^\beta_b}\partial_\alpha\tensor{\Lambda}{^a_\beta}$. We also remark, that since the metric is locally flat, there is no real distinction between lower and upper indices.

Since $\tensor{\omega}{_\alpha^a_b}=\Omega_\alpha\tensor{\varepsilon}{^a_b}$, $\tensor{\omega}{_\alpha^a_b}q^b\mathbf{E}_a=\Omega_\alpha \tensor{\varepsilon}{^a_b} q^b\mathbf{E}_a=\Omega_\alpha\left(q^2\mathbf{E}_1-q^1\mathbf{E}_2\right)$. However, since $\mathbf{N}=\mathbf{E}_1\times\mathbf{E}_2$, we have $\mathbf{N}\times\mathbf{E}_1=\mathbf{E}_2$ and $\mathbf{N}\times\mathbf{E}_2=-\mathbf{E}_1$, which gives $q^2\mathbf{E}_1-q^1\mathbf{E}_2=-q^1(\mathbf{N}\times\mathbf{E}_1)-q^2(\mathbf{N}\times\mathbf{E}_2)=-\mathbf{N}\times\mathbf{q}$. Therefore, 
\begin{equation}
    \nabla_\alpha\mathbf{q} = \partial_\alpha\mathbf{q} - \Omega_\alpha \mathbf{N}\times\mathbf{q}.
\end{equation}
The spin connection, thus, describes how the orthonormal basis rotates around the surface normal as one moves along the surface. If the vector field $\mathbf{q}$ is defined along the curve $\boldsymbol{\gamma}$ parametrised by $s$, we have \begin{eqnarray}
\frac{\mathrm{D}\mathbf{q}}{\mathrm{d}s} &=& \dot{\gamma}^\alpha \nabla_\alpha \mathbf{q}\nonumber\\
&=& \dot{\gamma}^\alpha \partial_\alpha \mathbf{q} - \dot{\gamma}^\alpha \Omega_\alpha \mathbf{N} \times \mathbf{q}\nonumber\\
&=& \frac{\mathrm{d}\mathbf{q}}{\mathrm{d}s} - \Omega_s\mathbf{N} \times \mathbf{q},    
\end{eqnarray} 
where $\Omega_s = \dot{\gamma}^\alpha \Omega_\alpha$. We can, therefore, think of $\Omega_s\mathbf{N}$ as the instantaneous angular velocity due to geometric torque applied to the vector $\mathbf{q}$ as it is transported along the curve $\boldsymbol{\gamma}$.

Now let's consider rotation of the frame basis around the local normal vector by an angle $\theta$. For a vector $\mathbf{q}=q^a\mathbf{E}_a$, we have $q'^a = \tensor{R}{^a_b}q^b$ and $\mathbf{E}'_a=\tensor{R}{_a^b}\mathbf{E}_b=\tensor{(R^{-1})}{^b_a}\mathbf{E}_b$, where $\tensor{R}{^a_b}$ are elements of the rotation matrix. We require $\left(\nabla_\alpha\mathbf{q}\right)'^a=\tensor{R}{^a_b}\left(\nabla_\alpha\mathbf{q}\right)^b$. Therefore, 
\begin{eqnarray}
 \left(\nabla_\alpha\mathbf{q}\right)'^a &=& \partial_\alpha q'^a + \tensor{{\omega'}}{_\alpha^a_b}q'^b \nonumber\\
 &=&\partial_\alpha(\tensor{R}{^a_c}q^c) + \tensor{{\omega'}}{_\alpha^a_b}\tensor{R}{^b_c}q^c\nonumber\\
 &=&\tensor{R}{^a_c}\partial_\alpha q^c + q^c\partial_\alpha\tensor{R}{^a_c} + \tensor{{\omega'}}{_\alpha^a_b}\tensor{R}{^b_c}q^c\nonumber\\
 &=&\tensor{R}{^a_b}\left(\nabla_\alpha\mathbf{q}\right)^b\nonumber\\
 &=&\tensor{R}{^a_b}\partial_\alpha q^b + \tensor{R}{^a_b}\tensor{\omega}{_\alpha^b_c}q^c.    
\end{eqnarray}
Matching coefficients of $q^c$ gives $\tensor{{\omega'}}{_\alpha^a_b}\tensor{R}{^b_c}=-\partial_\alpha\tensor{R}{^a_c} + \tensor{R}{^a_b}\tensor{\omega}{_\alpha^b_c}$. Multiplying both sides by $\tensor{(R^{-1})}{^c_d}$ gives
\begin{eqnarray}
\tensor{{\omega'}}{_\alpha^a_b}\tensor{R}{^b_c}\tensor{(R^{-1})}{^c_d}=&-&\tensor{(R^{-1})}{^c_d}\partial_\alpha\tensor{R}{^a_c}\nonumber\\
&+& \tensor{R}{^a_b}\tensor{\omega}{_\alpha^b_c}\tensor{(R^{-1})}{^c_d},
\end{eqnarray}
or 
\begin{eqnarray}
\tensor{{\omega'}}{_\alpha^a_b}\delta^b_d=\tensor{{\omega'}}{_\alpha^a_d}=&&\tensor{R}{^c_d}\partial_\alpha\tensor{(R^{-1})}{^a_c}\nonumber\\
&+& \tensor{R}{^a_b}\tensor{\omega}{_\alpha^b_c}\tensor{(R^{-1})}{^c_d}.    
\end{eqnarray}
We used $\tensor{R}{^c_d}\partial_\alpha\tensor{(R^{-1})}{^a_c}=-\tensor{(R^{-1})}{^c_d}\partial_\alpha\tensor{R}{^a_c}$, which is a direct consequence of $\tensor{R}{^a_c}\tensor{(R^{-1})}{^c_b}=\delta^a_b$.

In two dimensions, the rotation matrix is $\hat{R}=\big(\begin{smallmatrix}
  \cos\theta & -\sin\theta\\
  \sin\theta & \cos\theta
\end{smallmatrix}\big)$ and  $\hat{R}^{-1}=\big(\begin{smallmatrix}
  \cos\theta & \sin\theta\\
  -\sin\theta & \cos\theta
\end{smallmatrix}\big)$. Direct calculation shows that $\tensor{R}{^c_d}\partial_\alpha\tensor{(R^{-1})}{^a_c}=\partial_\alpha\theta\tensor{\varepsilon}{^a_d}$. The Levi-Civita symbol does not depened on the choice of coordinates, i.e.\ $\tensor{{\varepsilon'}}{^a_d}=\tensor{\varepsilon}{^a_d}$. Therefore, 
\begin{eqnarray}
\Omega'_\alpha\tensor{\varepsilon}{^a_b}&=&\Omega'_\alpha\tensor{{\varepsilon'}}{^a_b}\nonumber\\
&=&\tensor{{\omega'}}{_\alpha^a_b}\nonumber\\
&=&\tensor{R}{^c_b}\partial_\alpha\tensor{(R^{-1})}{^a_c} + \tensor{R}{^a_d}\Omega_\alpha \tensor{\varepsilon}{^d_c}\tensor{(R^{-1})}{^c_b}\nonumber\\
&=&\partial_\alpha\theta\tensor{\varepsilon}{^a_b}+\Omega_\alpha\tensor{\varepsilon}{^a_b}.    
\end{eqnarray}
We readily read off, $\Omega'_\alpha = \Omega_\alpha + \partial_\alpha\theta$. Therefore, rotating the frame around the normal introduces a pure gauge $\partial_\alpha\theta$ term to $\Omega_\alpha$.

By directly applying transformation properties for compoenents of vectors and the spin connection, it is strainghtforward to show that $\frac{\mathrm{D}q'^a}{\mathrm{d}s}=\tensor{R}{^a_b}\frac{\mathrm{D}q^b}{\mathrm{d}s}$, i.e.\ components of the derivative of $\mathbf{q}$ along the curve $\boldsymbol{\gamma}(s)$ transform the same way as the components of $\mathbf{q}$. Therefore, rotation of the frame basis around the local surface normal does not affect the equation of motion for $\mathbf{q}$.

\section{Geodesics on a subclass of axisymmetric surfaces}
\label{app:D}
We consider a subclass of axisymmetric surfaces that can be parametrised as $\mathbf{r}(r,\theta)=(r\cos\theta, r\sin\theta,h(r))$, where $h(r)$ is a smooth function of $r$ such that the derivative of $h$ with respect to $r$, $|h'(r)|<\infty$ everywhere. Coordinate $r$ labels meridians and coordinate $\theta$ labels parallels. The surface is obtained by rotating the curve $z=h(r)$ around the $z-$axis. Therefore, such surfaces are a subclass of surfaces of revolution. Coordinate tangent vectors are $\mathbf{e}_r = \partial_r\mathbf{r} = (\cos\theta,\sin\theta,h'(r))$ and $\mathbf{e}_\theta = \partial_\theta\mathbf{r} = (-r\sin\theta, r\cos\theta, 0)$. The surface normal is $\mathbf{N}=(1+h'(r)^2)^{-\frac{1}{2}}\left(-\cos (\theta ) h'(r),-\sin (\theta ) h'(r),1\right)$. We can readily find the components of the metric tensor $\hat{g}=\big(\begin{smallmatrix}
    1+h'^2 & 0 \\
    0 & r^2
\end{smallmatrix}\big)$. There are only four non-zero Christoffel symbols, $\tensor{\Gamma}{^r_{rr}} = \frac{h'h''}{1+h'^2}$, $\tensor{\Gamma}{^r_{\theta\theta}} = -\frac{r}{1+h'^2}$, and $\tensor{\Gamma}{^\theta_{r\theta}} = \tensor{\Gamma}{^\theta_{\theta r}}  = \frac{1}{r}$. Since $\mathbf{e}_r$ and $\mathbf{e}_\theta$ are everywhere orthogonal (but not orthonormal), it is straightforward to show that the covector associated with the spin connection has components $\Omega_r = 0$ and $\Omega_\theta = -\frac{1}{\sqrt{1 + h'^2}}$. $\Omega_r=0$ reflects the fact that the frame basis does not rotate as one moves along a meridian.

Using Christoffel symbols, the geodesic equations are
\begin{subequations}
    \begin{eqnarray}
        \ddot{r} + \frac{(\partial_r h)(\partial_r^2h) \dot{r}^2-r \dot{\theta}^2}{1 + h'^2}&=0 \label{eq:r-axi}\\
        \ddot{\theta} + \frac{2 \dot{r} \dot{\theta}}{r}&=0, \label{eq:theta-axi}
    \end{eqnarray}
\end{subequations}
and, for clarity, we denoted derivatives with respect to the curve parameter with an overdot and derivatives with respect to $r$ as $\partial_r$. These equations have two first integrals, which can be found by direct manipulation. Here, instead, we take a more physical approach. 

We recall that geodesics minimise length between two points. However, finding a curve that minimises the integral of $\mathrm{d}s^2$, where $\mathrm{d}s$ is the infinitesimal arc length along the curve, is simpler. We recall that $\mathrm{d}s^2 = g_{\alpha\beta}\mathrm{d}u^\alpha\mathrm{d}u^\beta=(1+h'(r)^2)\mathrm{d}r^2 + r^2\mathrm{d}\theta^2$. Since, $\mathrm{d}s$ is defined along a curve, $r$ and $\theta$ are functions of $t$, and we have $\mathrm{d}s^2=\left((1+h'(r)^2)\dot{r}^2 + r^2\dot{\theta}^2\right)\mathrm{d}t^2$. We, therefore, seek to minimise the functional $S=\int_{t_0}^{t_1}\mathcal{L}(\dot{r},\dot{\theta},r)\mathrm{d}t$, with the geodesic Lagrangian
\begin{equation}
  \mathcal{L}(\dot{r},\dot{\theta},r) =  \frac{1}{2}\left((1+h'(r)^2)\dot{r}^2 + r^2\dot{\theta}^2\right), 
\end{equation}
where we added the factor of $1/2$ since it makes the expression resemble the expression for the kinetic energy of a particle expressed in polar coordinates, albeit with a somewhat peculiar prefactor of the $\dot{r}^2$ term. Since $\mathcal{L}$ does not depend on $\theta$, we have $\frac{\partial\mathcal{L}}{\partial\theta}=0$, which gives $\frac{\mathrm{d}}{\mathrm{d}t}\left(\frac{\partial\mathcal{L}}{\partial\dot{\theta}}\right)=0$, or
\begin{equation}
    \frac{\partial\mathcal{L}}{\partial\dot{\theta}} = r^2\dot{\theta} = \ell.
\end{equation}
Therefore, $\ell$ is the angular momentum conserved along a geodesic. This is unsurprising since an active particle follows a geodesic without applied torques, and in the absence of torque, the angular momentum is conserved. 

If we recall that the conjugate momentum $p_\theta = \frac{\partial\mathcal{L}}{\partial\dot{\theta}}=r^2\dot{\theta}=\ell$, we readily see that we can write the energy (or the Hamiltonian),
\begin{equation}
    E = \frac{1}{2}\left((1+h'(r)^2)\dot{r}^2 + \frac{\ell^2}{r^2}\right),
\end{equation}
which is another first integral. The velocity vector along the geodesic $\mathbf{v}=\dot{r}\mathbf{e}_r + \dot{\theta}\mathbf{e}_\theta$, so $\mathbf{v}\cdot\mathbf{v} = (1+(\partial_r h)^2)\dot{r}^2 + r^2\dot{\theta}^2 = (1+(\partial_r h)^2)\dot{r}^2 + \frac{\ell^2}{r^2} = E$. Therefore, $E = |\mathbf{v}|^2$. Since the speed along a geodesic is constant, we can always reparametrise it such that $|\mathbf{v}|=1$, and, hence, $E=1$ without loss of generality. The equation for $\dot{r}$ this, therefore,
\begin{equation}
    \label{eq:r-dot-final}
    \dot{r}^2 = \frac{2}{1+(\partial_r h)^2}\left(1-\frac{\ell^2}{2r^2}\right).
\end{equation}
We can map the last equation to the radial motion of a particle in a central potential where $\frac{1}{2}(1+(\partial_r h)^2)\dot{r}$ is analogue to the radial kinetic energy with variable ``mass'' $1+(\partial_r h)^2$ and $\frac{\ell^2}{2r^2}$ is the effective (``centrifugal'') potential. 

In the case $\ell = 0$, the right hand side in Eq.\ \eqref{eq:r-dot-final}, $2/(1+(\partial_r h)^2)>0$. Geodesics are meridian that pass through the origin and continue to infinity. For $\ell\neq0$, the geodesic reaches $r_\mathrm{min}=|\ell|$ to the riding before being ``deflected'' to infinity. 

Finally, we consider the case $\dot{r}=0$, here the trajectory is along a parallel. A direct calculation shows that the geodesic curvature of a parallel with $r=R$ is $k_g = \left|\nabla_{\mathbf{t}}\mathbf{t}\right| = 1/(R\sqrt{1+h'(R)^2})$, where $\mathbf{t}$ is the unit-length tangent to the the parallel. We see that $k_g=0$ would require $h'(R)\to\infty$, which is contradictory to our assumption of $h'$ being finite everywhere. Therefore, there are no closed geodesics.

An in-depth discussion about geodesic on surfaces of revolution can be found in Ref.\ \cite{jantzen2012geodesics}.

\section{Jacobi fields}
\label{app:E}
Let $\boldsymbol{\gamma}_\mu(s)$ be a family of geodesics with $s$ being the arc length parameter of a geodesic and $\mu$ the parameter that labels different geodesics within the family. The Jacobi field is defined along the geodesic $\mu=0$ as $\boldsymbol{\mathcal{J}}(s)=\left.\frac{\partial\boldsymbol{\gamma}_\mu(s)}{\partial \mu}\right|_{\mu=0}$ and it describes how the nearby geodesics diverge or converge from it, i.e.\ it encodes how curvature affects the relative acceleration of geodesics \cite{frankel2011geometry}. Assuming that $\boldsymbol{\mathcal{J}}=\mathcal{J}\mathbf{l}$ ($\mathbf{l}$ is the tangent normal vector defined in Appendix \ref{app:B}), i.e.\ $\boldsymbol{\mathcal{J}}$ is orthogonal to the tangent vector, the Jacobi field satisfies the equation
\begin{equation}
    \frac{\mathrm{d}^2\mathcal{J}}{\mathrm{d}s^2} + K(s)\mathcal{J}(s) = 0,
\end{equation}
where $K$ is the Gaussian curvature along $\boldsymbol{\gamma}_0(s)$. If $K>0$ ($K<0$), nearby geodesics converge (diverge). For example, on a sphere of radius $R$, $K=1/R^2$ and $\mathcal{J}(s)=A\cos(s/R) + B\sin(r/R)$, where $A$ and $B$ are constants. So geodesics reconverge after a distance of $\pi R$, i.e.\ great circles either coincide or cross each other twice. Jacobi fields in general relativity characterise how nearby free-falling particles deviate from one another due to spacetime curvature, thus capturing the essence of gravitational tidal forces \cite{wald2010general}. 

\providecommand{\noopsort}[1]{}\providecommand{\singleletter}[1]{#1}%

\end{document}